# Noncommutative Nonlinear Supersymmetry

Hitoshi NISHINO[1] and Subhash RAJPOOT[2]

*Department of Physics & Astronomy*
*California State University*
*1250 Bellflower Boulevard*
*Long Beach, CA 90840*

**Abstract**

We present noncommutative nonlinear supersymmetric theories. The first example is a non-polynomial Akulov-Volkov-type lagrangian with noncommutative nonlinear global supersymmetry in arbitrary space-time dimensions. The second example is the generalization of this lagrangian to Dirac-Born-Infeld lagrangian with nonlinear supersymmetry realized in dimensions $D = 2, 3, 4, 6$ and $10$.

[1] E-Mail: hnishino@csulb.edu
[2] E-Mail: rajpoot@csulb.edu

# 1. Introduction

The importance of noncommutative geometry has been widely recognized, motivated by the recent developments of M-theory, [1] open superstrings [2], or D-branes [3][4][5] leading to noncommutative space-time coordinates [6][7]. In fact, the low energy effective theory of open strings attached to noncommutative branes becomes a noncommutative gauge theory [8]. Another example is a recent study [9] showing the equivalence between Dirac-Born-Infeld (DBI) theory with noncommutative gauge field strength and the ordinary DBI theory under so-called Seiberg-Witten map [7]. It has been also pointed out that type IIB matrix model with D-brane backgrounds can be interpreted as noncommutative Yang-Mills theory [10].

In noncommutative geometry, all the products in the theory are replaced by so-called $\star$ product involving the constant tensor $\theta^{\mu\nu}$ [11]. The next natural step to be considered is to make such algebra consistent with general covariance, which could possibly lead to the consistent formulation of noncommutative supergravity. However, there seems to be some fundamental problem with such trials, due to the difficulty of choosing the right measure, and/or dealing with complex metric with the right degrees of freedom [12][13]. There seems to be a persistent problem for unifying noncommutativity with the concept of metrics in gravity [12][13], not to mention supergravity with local supersymmetry.

As far as global supersymmetry is concerned, there has been considerable progress in noncommutative theories, *e.g.,* at quantum level [14], or in superspace with Moyal-Weyl deformations [11] for supersymmetric DBI theory [15]. Also ten-dimensional (10D) supersymmetric Yang-Mills has been generalized to noncommutative case, including the $F^4$-order corrections as the first non-trivial terms for supersymmetric DBI lagrangian [16] which can serve as an underlying theory of all noncommutative supersymmetric Yang-Mills theory in $D \leq 9$.

There has been, however, a different realization of global supersymmetry, called nonlinear realization. About three decades have past, since Volkov and Akulov (VA) gave a lagrangian for nonlinear supersymmetry in 4D in terms of Nambu-Goldstone fermion in 1972 [17]. Interestingly, it has been also known that nonlinear supersymmetries are not peculiar to 4D, but such formulations are universal in arbitrary space-time dimensions for both simple and extended supersymmetries [18]. In such a universal formulation, the lagrangian is given in terms of 'vielbein' as a generalization of the vierbein in the original VA lagrangian in 4D [17]. The inclusion of non-Abelian field strength as in DBI action is also shown to be straightforward [18].

Considering these long and recent developments as well, we realize the importance of combining the two concepts, *i.e.,* noncommutative algebra [6] and nonlinear supersymmetry [17][18]. Such a trial is also strongly motivated by D-brane physics [3][4] related to superstrings [2] and M-theory [1]. In fact, a typical example is the pioneering work on su-persymmetric DBI action in 10D by Aganagic, Popescu and Schwarz [5], and it is a natural next question whether such a lagrangian can be compatible with noncommutativity. In our present paper, we establish explicit lagrangians which are noncommutative generalization of DBI action with nonlinear supersymmetries.

As a preliminary step in the next section, we first present the noncommutative generalization of VA actions in arbitrary space-time dimensions. Based on this, we study noncommutative generalization of supersymmetric DBI action with nonlinear supersymmetry in dimensions $D = 2, 3, 4, 6$ and $10$ in section 3. Section 4 is for concluding remarks. Appendix A is devoted to the detailed explanations for flipping/hermiticity properties of fermions in general space-time dimensions $^\forall D$. Appendix B is for a lemma for general variations of noncommutative functionals. Appendix C is for a lemma related to the hermitian conjugation in (2.13).

## 2. Noncommutative VA Lagrangian in $^\forall D$

We first present our result, and subsequently we explain its notational or technical details. Our total action is valid in $^\forall D$ space-time dimensions with the usual signature $(\eta_{mn}) =$ diag. $(+, \overbrace{-, -, \cdots, -}^{D-1})$:

$$I_{\text{VA}} \equiv \int d^D x \, \text{sdet}_\star(E_\mu{}^m) \equiv \int d^D x \, E \tag{2.1a}$$

$$= \int d^D x \, [\,(-1)^{D-1} \text{sdet}_\star(g_{\mu\nu})\,]_\star^{1/2} \equiv \int d^D x \, \tilde{g}_\star^{1/2} \ , \tag{2.1b}$$

where $E_\mu{}^m$ is our vielbein and $g_{\mu\nu}$ is our 'metric':

$$E_\mu{}^m \equiv \delta_\mu{}^m + \mathcal{S}(i\overline{\lambda}\gamma^m \star \partial_\mu \lambda) \equiv \delta_\mu{}^m + \Lambda_\mu{}^m \ ,$$

$$g_{\mu\nu} \equiv \mathcal{S}(E_\mu{}^m \star \eta_{mn} \star E_\nu{}^n) \tag{2.2a}$$

$$= \mathcal{S}[\,\eta_{\mu\nu} + 2i(\overline{\lambda}\gamma_{(\mu} \star \partial_{\nu)}\lambda) - (\overline{\lambda} \star \gamma^m \partial_\mu \lambda) \star (\overline{\lambda} \star \gamma_m \partial_\nu \lambda)\,] = g_{\nu\mu} \ . \tag{2.2b}$$

The $\lambda$ is a (symplectic) (pseudo)Majorana spinor which is possible in any space-time dimension [19].[3] As usual, the symbol $\star$ refers to a noncommutative product defined typically in terms of two arbitrary fields $f(x)$ and $g(x)$ by [11]

$$f \star g \equiv f \, \exp\,(i\overleftarrow{\partial}_\mu \theta^{\mu\nu} \overrightarrow{\partial}_\nu)\, g \equiv \sum_{n=1}^\infty \frac{(+i)^n}{n!} \theta^{\mu_1 \nu_1} \cdots \theta^{\mu_n \nu_n} (\partial_{\mu_1} \cdots \partial_{\mu_n} f)(\partial_{\nu_1} \cdots \partial_{\nu_n} g) \ . \tag{2.3}$$

---

[3]The fermion $\lambda$ may carry implicit $Sp(1)$ indices, if it is symplectic (pseudo)Majorana spinor [19]. See Appendix A for more details.

Any subscript symbol $_\star$ therefore refers to expressions containing such $\star$ products, such as the determinants which are symmetrized by the symmetrization operator $\mathcal{S}$:

$$E \equiv \mathrm{sdet}_\star(E_\mu{}^m) \equiv \mathcal{S}\left[\tfrac{1}{D!} \epsilon^{\mu_1\cdots\mu_D}\epsilon_{m_1\cdots m_D} E_{\mu_1}{}^{m_1} \star \cdots \star E_{\mu_D}{}^{m_D}\right] , \tag{2.4a}$$

$$g \equiv \mathrm{sdet}_\star(g_{\mu\nu}) \equiv \mathcal{S}\left[\tfrac{1}{D!} \epsilon^{\mu_1\cdots\mu_D} \epsilon^{\nu_1\cdots\nu_D} G_{\mu_1\nu_1} \star \cdots \star G_{\mu_D\nu_D}\right] , \tag{2.4b}$$

$$\widetilde{g} \equiv (-1)^{D-1} g . \tag{2.4c}$$

The factor $(-1)^{D-1}$ is need in (2.1b), due to $\det(\eta_{mn}) = (-1)^{D-1}$. The $\mathcal{S}$-operation is the total symmetrization of any $\star$ product:

$$\mathcal{S}(A_1 \star \cdots \star A_n) \equiv \tfrac{1}{n!}\left[A_1 \star \cdots \star A_n + (n!-1)\text{-permutations}\right] , \tag{2.5}$$

where in the remaining $(n!-1)$ permutations, we have to take into account all the Grassmann parities of the fields $A_1, \cdots, A_n$.[4] Note that the effect of this $\mathcal{S}$-operation is not only symmetrizing the noncommutative product, but also extracting only the real part of the total expression, as can be easily confirmed starting with the definition (2.3). The symmetrization operation is needed in (2.2) for the metric to be symmetric. The fractional power such as $\widetilde{g}_\star^{1/2}$ can be consistently defined by the infinite series

$$(1+f)^p_\star \equiv \sum_{n=0}^\infty \frac{p(p-1)\cdots(p-n+1)}{n!} \overbrace{f \star \cdots \star f}^{n} \quad (p \in \mathbb{R}) . \tag{2.6}$$

Here $p$ can be any real number, not necessarily $1/2$ or $-1$. The split $1+f$ seems to be always needed, in such a way that this infinite series makes sense.

As explained in [18] for commutative case, the distinction between the two groups of indices $\mu, \nu, \cdots$ and $m, n, \cdots$ is 'formal', in order to use the analogy with general coordinate transformations. The meaning of this becomes clearer, when we proceed.

The inverse vielbein is defined again as an infinite series

$$E_m{}^\mu \equiv [(I+\Lambda)^{-1}_\star]_m{}^\mu \equiv (I - \Lambda + \Lambda \star \Lambda - \Lambda \star \Lambda \star \Lambda + \cdots)_m{}^\mu$$
$$\equiv (I - \Lambda + \Lambda^2_\star - \Lambda^3_\star + \cdots)_m{}^\mu \tag{2.7}$$

Defined in this fashion, $E_m{}^\mu$ is unique, satisfying the ortho-normality conditions

$$\mathcal{S}(E_\mu{}^m \star E_m{}^\nu) = E_\mu{}^m \star E_m{}^\nu = \delta_\mu{}^\nu , \quad \mathcal{S}(E_m{}^\mu \star E_\mu{}^n) = E_m{}^\mu \star E_\mu{}^n = \delta_m{}^n . \tag{2.8}$$

The inverse metric $G^{\mu\nu}$ is defined by

$$g^{\mu\nu} = \mathcal{S}(E_m{}^\mu \star \eta^{mn} \star E_n{}^\nu) , \tag{2.9}$$

---

[4] Since our motivation is to develop noncommutative version of VA action, our definition of the determinant itself contains noncommutativity. This point is slightly different from the commutative determinant used in DBI action in [7].

satisfying the conditions

$$\mathcal{S}(g_{\mu\nu} \star g^{\nu\rho}) = \delta_\mu{}^\rho \ , \qquad \mathcal{S}(g^{\mu\nu} \star g_{\nu\rho}) = \delta_\rho{}^\mu \ . \tag{2.10}$$

It is sometimes important to set up the complex-conjugate acting on the $\star$ products, as

$$(A \star B)^\dagger \equiv (B^\dagger) \star (A^\dagger) \ . \tag{2.11}$$

This rule is valid, even though there is a complex exponent with $\theta^{\mu\nu}$ implicitly in the $\star$ product, due to $\theta^{\mu\nu} = -\theta^{\nu\mu}$. Accordingly, our 'vielbein' defined by (2.2a) is real. To show this, we use the general hermiticity feature of (pseudo)Majorana spinors in $D$-dimensional space-time in the *commutative* case that

$$[\,i(\overline\lambda \gamma_\mu \partial_\nu \chi)\,]^\dagger = i(\overline\lambda \gamma_\mu \partial_\nu \chi) \ , \tag{2.12}$$

for the inner product of two (symplectic) (pseudo)Majorana spinors $\chi$ and $\lambda$.[5] Now in the noncommutative case, we can confirm the generalization of this with the inclusion of the $\mathcal{S}$-operator, as

$$\{\mathcal{S}[\,i(\overline\lambda \star \gamma_\mu \partial_\nu \chi)\,]\}^\dagger = \mathcal{S}[\,i(\overline\lambda \star \gamma_\mu \partial_\nu \chi)\,] \ . \tag{2.13}$$

Here the details of the confirmation is given in Appendix C. By supplying the $Sp(1)$-indices $A, B, \cdots = 1, 2$, the proof goes in a parallel way for the symplectic (pseudo)Majorana case, too. Now the hermiticity of $E_\mu{}^m$ is transparent, because $(\Lambda_\mu{}^m)^\dagger = \Lambda_\mu{}^m$.

Our action $I_{\rm VA}$ (2.1) is invariant under global nonlinear supersymmetry

$$\delta_Q \lambda = \epsilon + \mathcal{S}[\,i(\overline\epsilon \gamma^\mu \lambda) \star \partial_\mu \lambda\,] \equiv \epsilon + \mathcal{S}(\xi^\mu \star \partial_\mu \lambda) \ , \qquad \xi^\mu \equiv i(\overline\epsilon \gamma^\mu \lambda) \ . \tag{2.14}$$

Relevantly, $E_\mu{}^m$ and $g_{\mu\nu}$ transform as

$$\delta_Q E_\mu{}^m = \mathcal{S}[\,\xi^\nu \star \partial_\nu E_\mu{}^m + (\partial_\mu \xi^\nu) \star E_\nu{}^m\,] \ , \tag{2.15a}$$

$$\delta_Q g_{\mu\nu} = \mathcal{S}[\,\xi^\rho \star \partial_\rho g_{\mu\nu} + (\partial_\mu \xi^\rho) \star g_{\rho\nu} + (\partial_\nu \xi^\rho) \star g_{\mu\rho}\,] \ . \tag{2.15b}$$

In other words, these fields are transforming formally the same as 'general coordinate transformations'.

Since our $E_\mu{}^m$ and $g_{\mu\nu}$ are transforming, as if they were under general coordinate transformations, the invariance confirmation of our action $I_{\rm AV}$ can be confirmed as follows: First, consider the variation

$$\delta E \equiv \delta[\,{\rm sdet}_\star(E_\mu{}^m)\,] = \mathcal{S}[\,E \star E_m{}^\mu \star (\delta E_\mu{}^m)\,] \ , \tag{2.16}$$

---

[5]These fermions can be symplectic (pseudo)Majorana spinors in some space-time dimensions. In such a case, we need additional $Sp(1)$ indices [19] (Cf. Appendix A).

for an arbitrary variation $\delta E_\mu{}^m$, confirmed as

$$\begin{aligned}
\text{(LHS)} &= \delta E = \delta\Big[\,\frac{1}{D!}\,\epsilon^{\mu_1\cdots\mu_D}\,\epsilon_{m_1\cdots m_D} E_{\mu_1}{}^{m_1}\star\cdots\star E_{\mu_D}{}^{m_D}\,\Big] \\
&= \frac{1}{(D-1)!}\,\epsilon^{\mu_1\cdots\mu_D}\,\epsilon_{m_1\cdots m_D}\mathcal{S}[\,(\delta E_{\mu_1}{}^{m_1})\star E_{\mu_2}{}^{m_2}\star\cdots\star E_{\mu_D}{}^{m_D}\,] \\
&= \frac{1}{(D-1)!}\,\epsilon_{m_1\cdots m_D}\,\epsilon^{n_1 r_2\cdots r_D}\mathcal{S}[\,E\star E_{n_1}{}^{\mu_1}\star E_{r_2}{}^{\rho_2}\star\cdots\star E_{r_D}{}^{\rho_D}\star(\delta E_{\mu_1}{}^{m_1}) \\
&\qquad\qquad\qquad\qquad \star E_{\rho_2}{}^{m_2}\star\cdots\star E_{\rho_D}{}^{m_D}\,] \\
&= \frac{1}{(D-1)!}\,\epsilon_{m_1\cdots m_D}\,\epsilon^{n_1 m_2\cdots m_D}\mathcal{S}[\,E\star E_{n_1}{}^{\mu_1}\star(\delta E_{\mu_1}{}^{m_1})\,] \\
&= \mathcal{S}[\,E\star E_m{}^\mu\star(\delta E_\mu{}^m)\,] = \text{(RHS)}\ . \qquad (2.17)
\end{aligned}$$

Here use is made of the relationships (2.8), $\epsilon^{m r_1\cdots r_{D-1}}\,\epsilon^n{}_{r_1\cdots r_{D-1}} = (-1)^{D-1}(D-1)!$, and

$$\epsilon^{\mu_1\cdots\mu_D} = \mathcal{S}(\epsilon^{m_1\cdots m_D} E\star E_{m_1}{}^{\mu_1}\star\cdots\star E_{m_D}{}^{\mu_D})\ , \qquad (2.18)$$

which in turn is confirmed as

$$\begin{aligned}
\text{(RHS)} &= \mathcal{S}[\,\epsilon^{m_1\cdots m_D} E\star E_{m_1}{}^{\mu_1}\star\cdots\star E_{m_D}{}^{\mu_D}\,] \\
&= \mathcal{S}[\,\frac{1}{D!}\,\epsilon^{\nu_1\cdots\nu_D}\,\epsilon_{n_1\cdots n_D}\epsilon^{m_1\cdots m_D} E_{\nu_1}{}^{n_1}\star\cdots\star E_{\nu_D}{}^{n_D}\star E_{[m_1}{}^{\mu_1}\star\cdots\star E_{m_D]}{}^{\mu_D}\,] \\
&= \mathcal{S}[\,\epsilon^{\nu_1\cdots\nu_D} E_{\nu_1}{}^{m_1}\star\cdots\star E_{\nu_D}{}^{m_D}\star E_{m_1}{}^{[\mu_1}\star\cdots\star E_{m_D}{}^{\mu_D]}\,] \\
&= \mathcal{S}[\,\epsilon^{\nu_1\cdots\nu_D}(E_{\nu_1}{}^{m_1}\star E_{m_1}{}^{[\mu_1|})\star(E_{\nu_2}{}^{m_2}\star E_{m_2}{}^{|\mu_2|})\star\cdots\star(E_{\nu_D}{}^{m_D}\star E_{m_D}{}^{|\mu_D]})\,] \\
&= \mathcal{S}[\,\epsilon^{\nu_1\cdots\nu_D}\delta_{\nu_1}{}^{[\mu_1}\delta_{\nu_2}{}^{\mu_2}\cdots\delta_{\nu_D}{}^{\mu_D]}\,] \\
&= \epsilon^{\mu_1\cdots\mu_D} = \text{(LHS)}\ . \qquad (2.19)
\end{aligned}$$

Second, using the key equation (2.17), we can confirm

$$\begin{aligned}
\delta_Q E &= \mathcal{S}[\,E\star E_m{}^\mu\star\{\xi^\nu\star\partial_\nu E_\mu{}^m + (\partial_\mu\xi^\nu)\star E_\nu{}^m\}\,] \\
&= \mathcal{S}[\,\xi^\nu\star\partial_\nu E + E\star E_m{}^\mu\star E_\nu{}^m\star\partial_\mu\xi^\nu\,] \\
&= \mathcal{S}(\xi^\mu\star\partial_\mu E + E\star\partial_\mu\xi^\mu) \\
&= \mathcal{S}[\,\partial_\mu(\xi^\mu\star E)\,] = \partial_\mu[\,\mathcal{S}(\xi^\mu\star E)\,] = \text{(total div.)}\ , \qquad (2.20)
\end{aligned}$$

leading to $\delta_Q I_{\text{VA}} = 0$.

We next establish the equivalence of (2.1a) to (2.1b). To this end, we prove the lemma

$$\mathcal{S}[\,(\text{sdet}_\star A)\star(\text{sdet}_\star B)\,] = \text{sdet}_\star(A\star B) \qquad (2.21)$$

for arbitrary $D\times D$ matrices $A_i{}^j$ and $B_j{}^k$, as

$$\begin{aligned}
\text{(LHS)} &= \mathcal{S}[\,\frac{1}{(D!)^2}\,\epsilon^{i_1\cdots i_D}\epsilon_{j_1\cdots j_D}\epsilon^{k_1\cdots k_D}\epsilon_{l_1\cdots l_D}(A_{i_1}{}^{j_1}\star\cdots\star A_{i_D}{}^{j_D})\star(A_{k_1}{}^{l_1}\star\cdots\star A_{k_D}{}^{l_D})\,] \\
&= \mathcal{S}[\,\frac{1}{D!}\,\epsilon^{i_1\cdots i_D}\epsilon_{l_1\cdots l_D}(A_{i_1}{}^{j_1}\star B_{j_1}{}^{l_1})\star\cdots\star(A_{i_D}{}^{j_D}\star B_{j_D}{}^{l_D})\,] \\
&= \mathcal{S}[\,\frac{1}{D!}\,\epsilon^{i_1\cdots i_D}\epsilon_{j_1\cdots j_D}(A\star B)_{i_1}{}^{j_1}\star\cdots\star(A\star B)_{i_D}{}^{j_D}\,] \\
&= \text{sdet}_\star(A\star B) = \text{(RHS)}\ , \qquad (2.22)
\end{aligned}$$

where we used the indices $i, j, \cdots$, because it is common to the indices $\mu, \nu, \cdots$ and $m, n, \cdots$.

Using (2.21), we can prove the relationship

$$\widetilde{g} \equiv (-1)^{D-1} g \equiv (-1)^{D-1} \operatorname{sdet}_\star(g_{\mu\nu}) = E_\star^2 ~, \tag{2.23}$$

as

$$\begin{aligned}
\text{(LHS)} &= (-1)^{D-1} \operatorname{sdet}_\star(g_{\mu\nu}) = (-1)^{D-1} \operatorname{sdet}_\star[\,\mathcal{S}(E_\mu{}^m \star \eta_{mn} \star E_\nu{}^n)\,] \\
&= (-1)^{D-1} \operatorname{sdet}_\star(E_\mu{}^m \star \eta_{mn} \star E_\nu{}^n) \\
&= (-1)^{D-1} [\,\operatorname{sdet}_\star(E_\mu{}^m)\,] \star [\,\operatorname{sdet}_\star(\eta_{mn})\,] \star [\,\operatorname{sdet}_\star(E_\nu{}^n)^T\,] \\
&= +[\,\operatorname{sdet}_\star(E_\mu{}^m)\,] \star [\,\operatorname{sdet}_\star(E_\nu{}^n)\,] \\
&= E \star E = E_\star^2 = \text{(RHS)} ~.
\end{aligned} \tag{2.24}$$

Even though (2.23) does not necessarily imply the equality $\widetilde{g}_\star^{1/2} = E$, it can be confirmed by help of the following lemma:

$$(A_\star^p) \star (A_\star^q) = A_\star^{p+q} \qquad (p,\, q \in \mathbb{R}) ~, \tag{2.25}$$

where an arbitrary real scalar $A$ itself can contain some $\star$ products in it. The lemma (2.25) can be confirmed by splitting $A \equiv 1 + a$, and

$$\begin{aligned}
\text{(LHS of (2.25))} &= A_\star^p \star A_\star^q = (1+a)_\star^p \star (1+a)_\star^q \\
&= \Big[ \sum_{n=0}^{\infty} \frac{p(p-1)(p-2)\cdots(p-n+1)}{n!} \overbrace{a \star \cdots \star a}^{n} \Big] \star \Big[ \sum_{m=0}^{\infty} \frac{q(q-1)(q-2)\cdots(q-m+1)}{n!} \overbrace{a \star \cdots \star a}^{m} \Big] \\
&= \sum_{m=0}^{\infty} \sum_{n=0}^{\infty} \frac{p(p-1)\cdots(p-n+1)}{n!} \frac{p(q-1)\cdots(q-m+1)}{m!} \overbrace{a \star \cdots \star a}^{m+n} \\
&= \sum_{N=0}^{\infty} \sum_{n=0}^{N} \frac{p(p-1)\cdots(p-n+1)}{n!} \frac{p(q-1)\cdots(q-N+n+1)}{(N-n)!} \overbrace{a \star \cdots \star a}^{N} \qquad (N \equiv m+n) ~.
\end{aligned} \tag{2.26}$$

We now use the identity for the usual commutative product $(1+a)^p (1+a)^q = (1+a)^{p+q}$:

$$\sum_{n=0}^{N} \frac{p(p-1)\cdots(p-n+1)}{n!} \frac{q(q-1)\cdots(q-N+n+1)}{(N-n)!} \equiv \frac{(p+q)(p+q-1)\cdots(p+q-N+1)}{N!} ~, \tag{2.27}$$

which simplifies (2.26), as

$$\begin{aligned}
\text{(LHS of (2.25))} &= \sum_{N=0}^{\infty} \frac{(p+q)(p+q-1)\cdots(p+q-N+1)}{N!} \overbrace{a \star \cdots \star a}^{N} \\
&= (1+a)_\star^{p+q} = A_\star^{p+q} = \text{(RHS of (2.25))} ~.
\end{aligned} \tag{2.28}$$

Once (2.25) is established, it is clear that

$$\widetilde{g}_\star^{1/2} = E \ , \tag{2.29}$$

because it satisfies

$$\widetilde{g}_\star^{1/2} \star \widetilde{g}_\star^{1/2} = \widetilde{g}_\star^1 = \widetilde{g} = E \star E \ , \tag{2.30}$$

by (2.25), as desired. As a corollary, the reality of the integrand $\widetilde{g}_\star^{1/2}$ in (2.1b) is easily seen.

The relationship (2.29) also provides an alternative confirmation of the invariance $\delta_Q I_{\rm VA}$, via $\delta_Q G_{\mu\nu}$ instead of $\delta_Q E_\mu{}^m$. First, note the lemma for an arbitrary variation $\delta$:

$$\delta[\,(F[\varphi])_\star^p\,] = p\mathcal{S}[\,(\delta F[\varphi]) \star (F[\varphi])_\star^{p-1}\,] \qquad (p \in \mathbb{R}) \ , \tag{2.31}$$

confirmed by the lemma (B.3) in Appendix B for a general variation of a noncommutative functional of $\varphi$. Second, we use the relationship

$$\delta\widetilde{g} = \mathcal{S}[\,\widetilde{g} \star (\delta g_{\mu\nu}) \star g^{\mu\nu}\,] \ , \tag{2.32}$$

confirmed as

$$\begin{aligned}
({\rm LHS}) &= \delta\widetilde{g} = \delta\Big[\,\frac{(-1)^{D-1}}{D!} \epsilon^{\mu_1\cdots\mu_D} \epsilon^{\nu_1\cdots\nu_D} \mathcal{S}(g_{\mu_1\nu_1} \star \cdots g_{\mu_D\nu_D})\,\Big] \\
&= \frac{(-1)^{D-1}}{(D-1)!} \epsilon^{\mu_1\cdots\mu_D} \epsilon^{\nu_1\cdots\nu_D} \mathcal{S}[\,(\delta g_{\mu_1\nu_1}) \star g_{\mu_2\nu_2} \star \cdots \star g_{\mu_D\nu_D}\,] \\
&= \frac{(-1)^{D-1}}{(D-1)!} \epsilon^{m_1\cdots m_D} \epsilon^{n_1\cdots n_D} \mathcal{S}[\,E \star E_{m_1}{}^{\mu_1} \star \cdots E_{m_D}{}^{\mu_D} \star (\delta g_{\mu_1\nu_1}) \\
&\qquad \star E \star E_{n_1}{}^{\nu_1} \star \cdots \star E_{n_D}{}^{\nu_D} \star g_{\mu_2\nu_2} \star \cdots \star g_{\mu_D\nu_D}\,] \\
&= \frac{(-1)^{D-1}}{(D-1)!} \epsilon^{m_1\cdots m_D} \epsilon^{n_1\cdots n_D} \mathcal{S}[\,E \star E \star (\delta g_{\mu_1\nu_1}) \star E_{m_1}{}^{\mu_1} \star E_{n_1}{}^{\nu_1} \\
&\qquad \star (E_{m_2}{}^{\mu_2} \star g_{\mu_2\nu_2} \star E_{n_2}{}^{\nu_2}) \star \cdots \star (E_{m_D}{}^{\mu_D} \star g_{\mu_D\nu_D} \star E_{n_D}{}^{\nu_D})\,] \\
&= \frac{(-1)^{D-1}}{(D-1)!} \epsilon^{m_1\cdots m_D} \epsilon^{n_1}{}_{m_2\cdots m_D} \mathcal{S}[\,E \star E \star (\delta g_{\mu_1\nu_1}) \star E_{m_1}{}^{\mu_1} \star E_{n_1}{}^{\nu_1}\,] \\
&= +\mathcal{S}[\,E \star E \star (\delta g_{\mu\nu}) \star E_m{}^\mu \star E^{m\nu}\,] \\
&= +\mathcal{S}[\,\widetilde{g} \star (\delta g_{\mu\nu}) \star g^{\mu\nu}\,] = ({\rm RHS}) \ . \tag{2.33}
\end{aligned}$$

Now $\delta_Q g_{\mu\nu}$ is obtained from (2.15a), as

$$\begin{aligned}
\delta_Q g_{\mu\nu} &= \delta_Q[\,\mathcal{S}(E_\mu{}^m \star E_{\nu m})\,] \\
&= \mathcal{S}[\,\{\xi^\rho \star \partial_\rho E_\mu{}^m + (\partial_\mu \xi^\rho) \star E_\rho{}^m\} \star E_{\nu m} \\
&\qquad + E_\mu{}^m \star \{\xi^\rho \star \partial_\rho E_{\nu m} + (\partial_\nu \xi^\rho) \star E_{\rho m}\}\,] \\
&= \mathcal{S}[\,\xi^\rho \star \partial_\rho(E_\mu{}^m \star E_{\nu m}) + (\partial_\mu \xi^\rho) \star E_\rho{}^m \star E_{\nu m} + (\partial_\nu \xi^\rho) \star E_\mu{}^m \star E_{\rho m}\,] \\
&= \mathcal{S}[\,\xi^\rho \star \partial_\rho g_{\mu\nu} + (\partial_\mu \xi^\rho) \star g_{\rho\nu} + (\partial_\nu \xi^\rho) \star g_{\mu\rho}\,] \ , \tag{2.34}
\end{aligned}$$

yielding (2.15b). The invariance of our action $I_{\text{VA}}$ can be now through $\delta_Q g_{\mu\nu}$ instead of $\delta_Q E_\mu{}^m$, as

$$\begin{aligned}
\delta_Q \widetilde{g}_\star^{1/2} &= +\tfrac{1}{2}\mathcal{S}[\,\widetilde{g}_\star^{-1/2} \star (\delta_Q \widetilde{g})\,] \\
&= +\tfrac{1}{2}\mathcal{S}[\,\widetilde{g}_\star^{-1/2} \star \widetilde{g} \star (\delta_Q g_{\mu\nu}) \star g^{\mu\nu}\,] \\
&= +\tfrac{1}{2}\mathcal{S}[\,\widetilde{g}_\star^{1/2} \star \{\xi^\rho \star (\partial_\rho g_{\mu\nu}) + 2(\partial_\mu \xi^\rho) \star g_{\rho\nu}\} \star g^{\mu\nu}\,] \\
&= +\tfrac{1}{2}\mathcal{S}[\,\widetilde{g}_\star^{1/2} \star \xi^\rho \star (\partial_\rho g_{\mu\nu}) \star g^{\mu\nu} + 2(\partial_\mu \xi^\rho) \star g_{\rho\nu} \star g^{\mu\nu}\,] \\
&= \mathcal{S}[\,\tfrac{1}{2}\widetilde{g}_\star^{-1/2} \star \xi^\rho \star (\partial_\rho \widetilde{g}) + \widetilde{g}_\star^{1/2} \star (\partial_\mu \xi^\mu)\,] \\
&= \mathcal{S}[\,(\partial_\mu \widetilde{g}_\star^{1/2}) \star \xi^\mu + \widetilde{g}_\star^{1/2} \star \partial_\mu \xi^\mu\,] \\
&= \partial_\mu[\,\mathcal{S}(\widetilde{g}_\star^{1/2} \star \xi^\mu)\,] = (\text{total div.}) \ .
\end{aligned} \qquad (2.35)$$

We emphasize that the noncommutative inverse matrix such as $E_m{}^\mu$, or noncommutative irrational functions such as $A_\star^p$ can be defined only in terms of infinite series as perturbations around unity, like (2.7) or (2.26). This is to avoid inconsistency that might arise, when these infinite series are expansions around non-unity numbers.

We have been relying on the transformation property of our vielbein or metric as (2.15) that simplifies the invariance confirmation $\delta_Q I_{\text{VA}} = 0$. As careful readers may have noticed, this may raise some question about the consistency with a 'constant' tensor $\theta^{\mu\nu}$. Because once we introduce 'general covariance' with 'curved' metric, such a constant tensor seems problematic.

However, the point here is that the transformation (2.15) is just a 'formality' used to simplify the computation, but there is no actual 'general covariance' in the system. Even *before* imposing noncommutativity with $\theta^{\mu\nu}$, we have already encountered this situation with the 'constant' tensor $\delta_\mu{}^m$, because this Kronecker's delta is to be literally constant, while we introduce transformations such as (2.15). Nevertheless, we know that this poses no problem, because transformation (2.15) is just a formality to use the analog with general coordinate transformation simplifying the invariance confirmation of the action. We know other examples such as the 'constant' vectorial parameter $\zeta^\mu \equiv 2i(\overline{\epsilon}_2 \gamma^\mu \epsilon_1)$ for translation arising out of the commutator of two supersymmetries. In the commutative case [17][18], we know that this 'constant' vector poses no problem for the same reason given above. As such, all the effect of constant $\theta^{\mu\nu}$ does not upset the basic structure of transformation (2.15) mimicking a 'general coordinate transformation'. Once this point is understood, we have no worry about the compatibility between the constant $\theta^{\mu\nu}$ and general covariance, because the latter is just a 'fake' symmetry of the system.

## 3. Noncommutative Supersymmetric DBI Lagrangian

Once we have understood a noncommutative generalization of VA lagrangian in $\forall D$ dimensions, it is relatively easy to generalize it to a DBI lagrangian [5] with nonlinear supersymmetry. The only caveat is that due to the Fierz arrangement involved for quartic fermion terms, the space-time dimensions will be restricted to be $D = 2, 3, 4, 6$ and $10$, as we will see shortly.

The generalization from the VA case occurs in the definition of the metric. Our action is now in terms of a new metric $G_{\mu\nu}$:

$$I_{\text{DBI}} \equiv \int d^D x \, [\,(-1)^{D-1} \text{sdet}_\star(G_{\mu\nu})\,]_\star^{1/2} \equiv \int d^D x \, \widetilde{G}_\star^{1/2} \;, \tag{3.1a}$$

$$\widetilde{G} \equiv (-1)^{D-1} G \equiv (-1)^{D-1} \text{sdet}_\star(G_{\mu\nu}) \;, \tag{3.1b}$$

where the previous metric (2.2) is now generalized to the new metric

$$G_{\mu\nu} \equiv \mathcal{S}[\,\eta_{\mu\nu} + 2i(\overline{\lambda} \star \gamma_\mu \partial_\nu \lambda) + F_{\mu\nu} - (\overline{\lambda} \star \gamma^m \partial_\mu \lambda)(\overline{\lambda} \star \gamma_m \partial_\nu \lambda)\,] \tag{3.2a}$$

$$\equiv \mathcal{S}[\,g_{\mu\nu} + 2i(\overline{\lambda} \star \gamma_{[\mu} \partial_{\nu]} \lambda) + F_{\mu\nu}\,] \;. \tag{3.2b}$$

Compared with $g_{\mu\nu}$, the difference is in the last two terms in (3.2b). A special case of this lagrangian in 10D corresponds to the lagrangian in [5]. The new field $A_\mu$ undergoes the supersymmetry transformation rule:

$$\delta_Q A_\mu = \mathcal{S}[\,\xi^\nu \star \partial_\nu A_\mu + (\partial_\mu \xi^\nu) \star A_\nu + \xi_\mu + \tfrac{i}{3}\xi^\rho \star (\overline{\lambda} \star \gamma_\rho \partial_\mu \lambda)\,] \;,$$

$$\delta_Q \lambda = \epsilon + \mathcal{S}[\,i(\overline{\epsilon}\gamma^\mu \lambda) \star \partial_\mu \lambda\,] \equiv \epsilon + \mathcal{S}(\xi^\mu \star \partial_\mu \lambda) \;, \tag{3.3}$$

where $\xi^\mu \equiv i(\overline{\epsilon}\gamma^\mu \lambda)$ is the same as the last section. This is a noncommutative and multi-dimensional generalization of the commutative case in 10D [5].

The invariance of our action $I_{\text{DBI}}$ can be confirmed in a way parallel to the previous case for $I_{\text{AV}}$, with the aid of the lemma

$$\delta_Q F_{\mu\nu} = \mathcal{S}[\,\xi^\rho \star \partial_\rho F_{\mu\nu} + (\partial_\mu \star \xi^\rho) \star F_{\rho\nu} + (\partial_\nu \xi^\rho) \star F_{\mu\rho} - 2i(\overline{\epsilon}\gamma_{[\mu}\partial_{\nu]}\lambda)$$
$$- \tfrac{2}{3}\partial_{[\mu|}[\,(\overline{\epsilon}\gamma^\rho \lambda) \star (\overline{\lambda}\gamma_\rho \partial_{|\nu]}\lambda)\,]\,] \;. \tag{3.4}$$

The confirmation of this lemma needs special care, associated with a Fierz rearrangement. This is because we need the equality

$$\mathcal{S}[\,(\overline{\epsilon}\gamma^\rho \partial_\mu \lambda) \star (\overline{\lambda}\gamma_\rho \partial_\nu \lambda)\,] - (\mu \leftrightarrow \nu) = \mathcal{S}[\,\tfrac{1}{3}\partial_\mu\{(\overline{\epsilon}\gamma^\rho \lambda) \star (\overline{\lambda}\gamma_\rho \partial_\nu \lambda)\}\,] - (\mu \leftrightarrow \nu) \;, \tag{3.5}$$

which is in turn confirmed by the Fierz identity[6]

$$(\gamma^m)_{(\underline{\alpha\beta}|}(\gamma_m)_{|\underline{\gamma})\underline{\delta}} \equiv 0 \;. \tag{3.6}$$

---
[6]Here the indices $\underline{\alpha}, \underline{\beta}, \underline{\gamma}, \underline{\delta}$ may contain also the $Sp(1)$ indices for symplectic (pseudo)Majorana spinors.

This identity holds only in space-time dimensions $D = 2, 3, 4, 6$ and 10 [20], so that the invariance of our action $I_{\text{DBI}}$ is valid only in these dimensions. The important ingredient here is the Fierz identity in its very universal form (3.6), which does not depend on the dimensionality of spinorial components.

The actual invariance confirmation of $I_{\text{DBI}}$ under (3.3) is parallel to that for (2.1), because the metric $G_{\mu\nu}$ transforms under (3.2) exactly as (2.15b):

$$\delta_Q G_{\mu\nu} = \mathcal{S}[\,\xi^\rho \star \partial_\rho G_{\mu\nu} + (\partial_\mu \xi^\rho) \star G_{\rho\nu} + (\partial_\nu \xi^\rho) \star G_{\mu\rho}\,] \quad. \tag{3.7}$$

Since the rest of the proof is parallel to that for the action (2.1), we will skip it here.

As an independent consistency check, we study the commutator of two supersymmetries on $A_\mu$:

$$[\delta_1, \delta_2] A_\mu = \zeta^\nu \partial_\nu A_\mu + \zeta_\mu \qquad (\zeta^\mu \equiv 2i(\bar{\epsilon}_2 \gamma^\mu \epsilon_1)) \quad. \tag{3.8}$$

In this computation, there arise four sorts of terms, (i) $\lambda^0$-terms, (ii) $\lambda^2$-terms, (iii) $\lambda^2 A$-terms, and (iv) $\lambda^4$-terms. The category (i) gives (3.8), while all others cancel themselves. One of the crucial identities we need is

$$\mathcal{S}[\,(\bar{\epsilon}_2 \gamma^\nu \lambda) \star (\bar{\epsilon}_1 \gamma_\nu \partial_\mu \lambda)\,] - {}_{(1\leftrightarrow 2)} = \mathcal{S}[\,-\tfrac{i}{4} \zeta^\nu (\bar{\lambda} \star \gamma_\nu \partial_\mu \lambda)\,] - {}_{(1\leftrightarrow 2)} \quad, \tag{3.9}$$

which is again valid only for the dimensions $D = 2, 3, 4, 6$ and 10, due to the Fierz identity (3.6). This provides an independent consistency check for our total system in these dimensions. As has been stated before, the existence of 'constant' vector $\zeta^\mu \equiv 2i(\bar{\epsilon}_2 \gamma^\mu \epsilon_1)$ poses no problem in our formulation, neither does the constant tensor $\theta^{\mu\nu}$. Needless to say, the result in [5] is a special case in 10D.

## 4. Concluding Remarks

In this paper, we have established the noncommutative version of VA lagrangian, and that of DBI lagrangian with nonlinear supersymmetry in space-time dimensions $2, 3, 4, 6$ and 10. The invariance of our actions under nonlinear supersymmetry has been confirmed by the use of various lemma, involving the symmetrized noncommutative determinants. The important new ingredient is that our noncommutative VA-type action is valid in $\forall D$, while our noncommutative DBI action is valid in $D = 2, 3, 4, 6$ and 10.

The difficulty of noncommutative generalization of supergravity has been well recognized for some time [12][21]. This is caused by the compatibility question between the enlarged complexified Lorentz symmetry such as $U(1,3)$ and spinor structure of such space-time manifolds. Even though our lagrangians have 'formal' metrics or vielbeins, we do not encounter such a problem, because $g_{\mu\nu}$, $G_{\mu\nu}$ or $E_\mu{}^m$ defined by (2.2) or (3.2) have the $\mathcal{S}$-operator which makes these fields real. Therefore no complexification of Lorentz symmetries, such as $U(1,3)$, is needed.

We have also seen that the existence of the 'constant' tensor $\theta^{\mu\nu}$ does not pose any problem with the 'general coordinate transformation' like (2.15) or (3.7). This is because general covariance is a 'fake' symmetry that does not actually exist in the system, but this is just for analogy that simplifies the computation for invariance confirmation. As a matter of fact, we have already encountered similar situations in the *commutative* case, such as the 'constant' Kronecker's delta $\delta_\mu{}^m$ in $E_\mu{}^m$, or the 'constant' vectorial parameter $\zeta^\mu \equiv 2i(\bar\epsilon_2 \gamma^\mu \epsilon_1)$ for the translation out of a supersymmetry commutator, none of which posed any problem [18].

Our action $I_{\text{DBI}}$ can be also regarded as the supersymmetric generalization of bosonic DBI action in dimensions $D = 2, 3, 4, 6$ and 10, in the same sense as the action in [5] is such a generalization in 10D. In other words, our actions are the generalization of VA action by vector fields, which in turn is a supersymmetric generalization of bosonic DBI action. The only caveat here is that supersymmetry in our system is realized nonlinearly, which is different from linear supersymmetry in terms of superfields formulated, *e.g*, in [15].

Since our actions are given as non-polynomial forms, the invariance under global supersymmetries is guaranteed to all orders in an expansion parameters, *e.g.*, the parameter $\alpha$ in $E_\mu{}^m = \delta_\mu{}^m + \alpha \Lambda_\mu{}^m$. In fact, the familiar $F^4$-term: $F_\mu{}^\nu F_\nu{}^\rho F_\rho{}^\sigma F_\sigma{}^\mu - (1/4)(F_{\mu\nu}^2)^2$ shows up among the lower-order terms in $I_{\text{DBI}}$.

Our noncommutative VA-type action is valid in $^\forall D$ space-time dimensions, while our noncommutative supersymmetric DBI action is valid in $D = 2, 3, 4, 6$ and 10. Thus it is similar to the restriction for linear global supersymmetries in $D \leq 10$, unless we sacrifice lorentz covariance [22]. This is in contrast to the commutative case of VA-type action formulated in $^\forall D$ dimensions [18].

We have seen that the invariance of our actions is by the intricate interplay among noncommutative determinants, noncommutative non-rational functions like square roots, and nonlinear supersymmetry transformation, all defined in terms of noncommutative products. The important key technique is the introduction of the symmetrization operator $\mathcal{S}$ that simplifies the whole computation drastically, making everything parallel to the proof in the *commutative* case.

## Appendix A: Flipping and Hermiticity Properties for Fermions

In this appendix, we analyze the flipping and hermiticity properties of fermionic bilinears, as promised in sections two and three.

Consider the general $D$-dimensional space-time of dimension with the signature $(+,\overbrace{-,-,\cdots,-}^{D-1})$ with the Clifford algebra

$$\{\gamma_m, \gamma_n\} = +2\,\eta_{mn} = +2\,\mathrm{diag.}\,(+,\overbrace{-,-,\cdots,-}^{D-1})\ . \tag{A.1}$$

For general treatment of spinors, we follow [19], where the relevant equations are such as

$$\begin{aligned}
(\gamma_0)^\dagger &= +\gamma_0\ , \qquad (\gamma_i)^\dagger = -\gamma_i \quad (i = 1,\,2,\,\cdots,\,D-1)\ , \\
(\gamma_m)^\dagger &= A\gamma_m A^{-1}\ , \qquad A \equiv \gamma_0\ , \qquad A^\dagger = A\ , \\
(\gamma_m)^* &= \eta B\gamma_m B^{-1}\ , \qquad B \equiv (A^T)^{-1} C^{-1}\ , \qquad A \equiv (B^T)^{-1} C\ , \\
(\gamma_m)^T &= +\eta C\gamma_m C^{-1}\ , \qquad C^\dagger C = +I\ , \qquad C^T = \epsilon\eta C\ .
\end{aligned} \tag{A.2}$$

Here the matrix $B$ is related to the complex conjugation of fermions, and $C$ is for the usual charge conjugation, both in the same notation as in [19], while $\epsilon$ and $\eta$ are $\pm 1$, depending on the difference $D-2$ between the space-like and time-like coordinates. There are in total four cases: $D-2 = 1,2,8$ (mod 8), or $6,7,8$ (mod 8), or $4,5,6$ (mod 8), or $2,3,4$ (mod 8) [19], tabulated equivalently as

| D | $\epsilon$ | $\eta$ | Fermions |
|---|---|---|---|
| $2,3,4$ (mod 8) | $+1$ | $-1$ | Majorana |
| $1,2,8$ (mod 8) | $+1$ | $+1$ | Pseudo-Majorana |
| $6,7,8$ (mod 8) | $-1$ | $-1$ | Symplectic Majorana |
| $4,5,6$ (mod 8) | $-1$ | $+1$ | Symplectic Pseudo-Majorana |

In the case of 'symplectic (pseudo)Majorana' spinors, we have an additional $Sp(1)$ indices $A,\,B,\,\cdots = 1,\,2$ on these fermions.

We next study the flipping property

$$(\overline{\psi}\gamma^{m_1\cdots m_n}\chi) = -\epsilon\eta^{n+1}(-1)^{n(n-1)/2}(\overline{\chi}\gamma^{m_1\cdots m_n}\psi)\ . \tag{A.3}$$

This can be proven by taking the transposition of the l.h.s., which is a scalar and intact under such an operation. As for symplectic (pseudo)-Majorana spinors, these includes also the $Sp(1)$ indices, e.g., the l.h.s. is $(\overline{\psi}^A\gamma^{m_1\cdots m_n}\chi^B)$, etc. Eq. (A.3) implies that for $D =$ 1, 2, 3, 4 and 8 (*mod* 8), we have the desirable antisymmetry $(\overline{\epsilon}_2 \gamma^m \epsilon_1) = -(\overline{\epsilon}_1 \gamma^m \epsilon_2)$. In the case of symplectic (pseudo)Majorana spinors in $D = 4, 5, 6, 7$ and 8 (mod 8), we need to multiply an extra $Sp(1)$ metric $(\epsilon_{AB}) \equiv \begin{pmatrix} 0 & +1 \\ -1 & 0 \end{pmatrix}$, like $(\overline{\epsilon}_2^A \gamma^m \epsilon_{1A}) \equiv (\overline{\epsilon}_2^A \gamma^m \epsilon_1^B) \epsilon_{BA} = -(\overline{\epsilon}_1^A \gamma^m \epsilon_{2A})$, as desired. Since these exhaust all the space-time dimensions, we have the desirable flipping properties needed for our VA-type or DBI-type actions in $^\forall D$.

The hermiticity operation for fermions are [19]

$$\overline{\psi} = \psi^\dagger A \ , \quad \overline{\psi}^\dagger = A\psi \ , \tag{A.4}$$

for (pseudo)Majorana spinors, so that we have

$$(\overline{\psi}\gamma^{m_1 \cdots m_n}\chi)^\dagger = -\epsilon \eta^{n+1}(\overline{\psi}\gamma^{m_1 \cdots m_n}\chi) \ , \tag{A.5}$$

while for symplectic (pseudo) Majorana spinors,

$$\overline{\psi}^A = \psi_A^\dagger A \ , \quad \overline{\psi}^{A\dagger} = A\psi_A \ , \tag{A.6}$$

so that we have

$$(\overline{\psi}^A \gamma^{m_1 \cdots m_n} \chi_B)^\dagger = -\epsilon \eta^{n+1}(\overline{\psi}_A \gamma^{m_1 \cdots m_n} \chi^B) \ . \tag{A.7}$$

The most important case is $n = 1$, which in turn implies the hermiticity of the combinations

$$\begin{cases} i(\overline{\lambda}\gamma_\mu \partial_\nu \lambda) \ , \ i(\overline{\epsilon}_2 \gamma^m \epsilon_1) & \text{for (pseudo)Majorana spinors } (\epsilon = +1) \ , \\ i(\overline{\lambda}^A \gamma_\mu \partial_\nu \lambda_A) \ , \ i(\overline{\epsilon}_2^A \gamma^m \epsilon_{1A}) & \text{for symplectic (pseudo)Majorana spinors } (\epsilon = -1) \ . \end{cases} \tag{A.8}$$

Accordingly, our $\Lambda_\mu{}^m$ is hermitian in all of these cases:

$$\Lambda_\mu{}^m \equiv \begin{cases} i(\overline{\lambda}\gamma^m \partial_\mu \lambda) & \text{for (pseudo)Majorana spinors } \ , \\ i(\overline{\lambda}^A \gamma^m \partial_\mu \lambda_A) & \text{for symplectic (pseudo)Majorana spinors } \ . \end{cases} \tag{A.9}$$

Needless to say, these cover any $(D-1) + 1$ space-time dimensions for $^\forall D$. This also verifies our statements associated with (2.13).

**Appendix B: Variation of Arbitrary Noncommutative Function of Fields**

We can prove a general lemma for a variation of the noncommutative generalization of a real functional. Suppose we have a real functional $H[\varphi]$ of a real field $\varphi$ defined by the Taylor expansion

$$H[\varphi] \equiv \sum_{n=0}^{\infty} \frac{a_n}{n!} \varphi^n \quad (a_n \in \mathbb{R}) \ . \tag{B.1}$$

Then a noncommutative generalization is

$$H_\star[\varphi] \equiv \sum_{n=0}^{\infty} \frac{a_n}{n!}\, \varphi_\star^n \quad . \tag{B.2}$$

We do not need an $\mathcal{S}$-operation here, because it is automatically symmetrized. Then the lemma we want to prove is

$$\delta\, H_\star[\varphi] = \mathcal{S}[\,H'_\star[\varphi] \star \delta\varphi\,] \quad . \tag{B.3}$$

Here the symbol $H'_\star[\varphi]$ implies the replacements of any product $\varphi^n$ in the definition of the derivative $H'[\varphi] \equiv dH[\varphi]/d\varphi$ in the commutative case by the noncommutative one $\varphi_\star^n$. This lemma is confirmed as

$$\begin{aligned}
(\text{LHS}) &= \sum_{n=0}^{\infty} \frac{a_n}{n!}\, \delta(\varphi_\star^n) = \sum_{n=0}^{\infty} \frac{a_n}{n!}\, \mathcal{S}[\,n\varphi_\star^{n-1} \star \delta\varphi\,] \\
&= \mathcal{S}\Big[\,\Big(\sum_{m=0}^{\infty} \frac{a_{m+1}}{m!}\, \varphi_\star^m\Big) \star \delta\varphi\,\Big] = \mathcal{S}[\,H'_\star[\varphi] \star \delta\varphi\,] = (\text{RHS}) \quad .
\end{aligned} \tag{B.4}$$

This lemma is general enough to cover the variations needed such as (2.31), when $H[\varphi] \equiv (F[\varphi])^p$.

### Appendix C: Confirmation of (2.13)

Here we give the detailed confirmation of (2.13). Note that our metric and vielbein are hermitian but not complex defined in a peculiar way with the $\mathcal{S}$-operator. Since this aspect was not covered in references in the past [12], it is better to demonstrate the details of its confirmation:

$$\begin{aligned}
(\text{LHS of (2.13)}) &= \{\mathcal{S}[\,i(\overline{\lambda} \star \gamma_\mu \partial_\nu \chi)\,]\}^\dagger = \mathcal{S}[\,\{i(\overline{\lambda} \star \gamma_\mu \partial_\nu \chi)\}^\dagger\,] \\
&= +\tfrac{1}{2}\Big[i(\overline{\lambda} \star \gamma_\mu \partial_\nu \chi) - i(\partial_\nu \overline{\chi}) \star \gamma_\mu \lambda\Big]^\dagger \\
&= +\tfrac{1}{2}\Big[+i \sum_{n=0}^{\infty} \frac{(+i)^n}{n!}\, \overline{\lambda} \overleftarrow{\partial}_{\rho_1} \cdots \overleftarrow{\partial}_{\rho_n} \gamma_\mu\, \theta^{\rho_1\sigma_1} \cdots \theta^{\rho_n\sigma_n} \partial_{\sigma_1} \cdots \partial_{\sigma_n} \partial_\nu \chi\Big]^\dagger \\
&\quad +\tfrac{1}{2}\Big[-i \sum_{n=0}^{\infty} \frac{(+i)^n}{n!}\, (\partial_\nu \overline{\chi}) \overleftarrow{\partial}_{\rho_1} \cdots \overleftarrow{\partial}_{\rho_n} \gamma_\mu\, \theta^{\rho_1\sigma_1} \cdots \theta^{\rho_n\sigma_n} \partial_{\sigma_1} \cdots \partial_{\sigma_n} \lambda\Big]^\dagger \\
&= +\tfrac{1}{2}(-i) \sum_{0}^{\infty} \frac{(-i)^n}{n!}\, (\partial_{\sigma_1} \cdots \partial_{\sigma_n} \chi^\dagger)(\gamma_\mu)^\dagger \theta^{\rho_1\sigma_1} \cdots \theta^{\rho_n\sigma_n} \partial_{\rho_1} \cdots \partial_{\rho_n} \overline{\lambda}^\dagger \\
&\quad +\tfrac{1}{2}(+i) \sum_{0}^{\infty} \frac{(-i)^n}{n!}\, (\partial_{\sigma_1} \cdots \partial_{\sigma_n} \lambda^\dagger)(\gamma_\mu)^\dagger \theta^{\rho_1\sigma_1} \cdots \theta^{\rho_n\sigma_n} \partial_{\rho_1} \cdots \partial_{\rho_n} \partial_\nu \overline{\chi}^\dagger \\
&= -\tfrac{i}{2} \sum_{0}^{\infty} \frac{(-i)^n}{n!}\, (\partial_{\sigma_1} \cdots \partial_{\sigma_n} \partial_\nu \overline{\chi} A^{-1})(A \gamma_\mu A^{-1})\, \theta^{\rho_1\sigma_1} \cdots \theta^{\rho_n\sigma_n} \partial_{\rho_1} \cdots \partial_{\rho_n} (A\lambda)
\end{aligned}$$

$$+ \frac{i}{2} \sum_0^\infty \frac{(-i)^n}{n!} (\partial_{\sigma_1} \cdots \partial_{\sigma_n} \overline{\lambda} A^{-1})(A\gamma_\mu A^{-1}) \theta^{\rho_1 \sigma_1} \cdots \theta^{\rho_n \sigma_n} \partial_{\rho_1} \cdots \partial_{\rho_n} \partial_\nu (A\chi)$$

$$= -\frac{i}{2} \sum_0^\infty \frac{(-i)^n}{n!} (\partial_{\sigma_1} \cdots \partial_{\sigma_n} \partial_\nu \overline{\chi}) \gamma_\mu \theta^{\rho_1 \sigma_1} \cdots \theta^{\rho_n \sigma_n} \partial_{\rho_1} \cdots \partial_{\rho_n} \lambda$$

$$+ \frac{i}{2} \sum_0^\infty \frac{(-i)^n}{n!} (\partial_{\sigma_1} \cdots \partial_{\sigma_n} \overline{\lambda}) \gamma_\mu \theta^{\rho_1 \sigma_1} \cdots \theta^{\rho_n \sigma_n} \partial_{\rho_1} \cdots \partial_{\rho_n} \partial_\nu \chi$$

$$= -\frac{i}{2} \sum_0^\infty \frac{(+i)^n}{n!} (\partial_{\sigma_1} \cdots \partial_{\sigma_n} \partial_\nu \overline{\chi}) \gamma_\mu \theta^{\sigma_1 \rho_1} \cdots \theta^{\sigma_n \rho_n} \partial_{\rho_1} \cdots \partial_{\rho_n} \lambda$$

$$+ \frac{i}{2} \sum_0^\infty \frac{(+i)^n}{n!} (\partial_{\sigma_1} \cdots \partial_{\sigma_n} \overline{\lambda}) \gamma_\mu \theta^{\sigma_1 \rho_1} \cdots \theta^{\sigma_n \rho_n} \partial_{\rho_1} \cdots \partial_{\rho_n} \partial_\nu \chi$$

$$= -\tfrac{i}{2}(\partial_\nu \overline{\chi}) \star \gamma_\mu \lambda + \tfrac{i}{2}(\overline{\lambda} \star \gamma_\mu \partial_\nu \chi) = \mathcal{S}[\, i(\overline{\lambda} \star \gamma_\mu \partial_\nu \chi)\,] = (\text{RHS of (2.13)}) \ . \qquad (\text{C.1})$$

Here use is also made of (A.2), and (A.4) for (pseudo)Majorana spinors $\lambda$ and $\chi$.

In the case of symplectic (pseudo)Majorana spinors, we can confirm

$$\{\mathcal{S}[\, i(\overline{\lambda}^A \star \gamma_\mu \partial_\nu \chi_A)\,]\}^\dagger = \mathcal{S}[\, i(\overline{\lambda}^A \star \gamma_\mu \partial_\nu \chi_A)\,] \ , \qquad (\text{C.2})$$

in a similar way. The special case $\chi = \lambda$ or $\chi_A = \lambda_A$ leads to our conclusion $(\Lambda_\mu{}^m)^\dagger = \Lambda_\mu{}^m$ as in section two.

This result is in a sense expected, because the $\mathcal{S}$-operation is effectively equivalent to adding the hermitian conjugate of the original expression. However, we emphasize that each step in (C.1) is the result of subtle interplay between flipping and hermiticity properties for (pseudo)Majorana spinors and $\gamma$-matrices.